\documentclass[12 pt]{article} 
\usepackage{times,bm}
\usepackage[affil-it,]{authblk}
\usepackage[pagebackref=false,colorlinks=true,linkcolor=blue,citecolor=blue,urlcolor=blue]{hyperref}
\usepackage{geometry}
\usepackage{graphicx}
\usepackage{cite}
\usepackage{amsmath}
\usepackage{amsfonts}
\usepackage{amssymb}
\usepackage{color}
\usepackage{float}
\usepackage{multirow,multicol}
\usepackage{tabulary}
\usepackage[flushleft]{threeparttable}
\usepackage{pgfplots,subfigure}
\usepackage{xcolor}
\numberwithin{equation}{section}
\usepackage{tikz}
\usepackage{ulem}
\usepackage{hyperref}
\usepackage{nameref}
\usepackage{comment}

\usepgflibrary{arrows}
\usetikzlibrary{shapes.callouts}
\tikzset{  
	level/.style   = { thick, },
	connect/.style = { dotted, red   },
	notice/.style  = { draw, rectangle callout, callout relative pointer={#1} },
	label/.style   = { text width=2cm }
}

\usepackage{letltxmacro}
\makeatletter
\let\oldr@@t\r@@t
\def\r@@t#1#2{%
	\setbox0=\hbox{$\oldr@@t#1{#2\,}$}\dimen0=\ht0
	\advance\dimen0-0.2\ht0
	\setbox2=\hbox{\vrule height\ht0 depth -\dimen0}%
	{\box0\lower0.4pt\box2}}
\LetLtxMacro{\oldsqrt}{\sqrt}
\renewcommand*{\sqrt}[2][\ ]{\oldsqrt[#1]{#2}}
\makeatother
\begin{document}

\newcommand{{\ri}}{{\rm{i}}}
\newcommand{{\Psibar}}{{\bar{\Psi}}}

\title{Relativistic Solutions of Generalized-Dunkl Harmonic and Anharmonic Oscillators} 


\author{\large \textit {S.\ Hassanabadi}$^{\ 1}$\footnote{s.hassanabadi@yahoo.com}~,~ \textit {J. Kříž}$^{\ 1}$\footnote{jan.kriz@uhk.cz}~,~\textit {B. C. Lütfüoğlu}$^{\ 1}$ \footnote{bekircanlutfuoglu@gmail.com(corresponding author)}~ and~\textit {H.\ Hassanabadi}$^{\ 1,\ 2}$\footnote{hha1349@gmail.com}  \\

\small \textit {$^{\ 1}$Department of Physics, University of Hradec Králové, Rokitanského 62, 500 03 Hradec Králové, Czechia}\\
\small \textit {$^{\ 3}$Faculty of Physics, Shahrood University of Technology, Shahrood, Iran}\\
}

\date{}
\maketitle

\begin{abstract}		
	Dunkl derivative enriches solutions by discussing parity due to its reflection operator. Very recently, one of the authors of this manuscript presented one of the most general forms of Dunkl derivative that depends on three Wigner parameters to have a better tuning. In this manuscript, we employ the latter generalized Dunkl derivative in a relativistic equation to examine two dimensional harmonic and anharmonic oscillators solutions. We obtain the solutions by Nikiforov-Uvarov and QES methods, respectively. We show that degenerate states can occur according to the Wigner parameter values.
	
\end{abstract}

\begin{small}	
Keywords: Generalized-Dunkl derivative; Klein-Gordon equation; harmonic potential energy, anharmonic potential energy, polar coordinates.	
\end{small}

\begin{small}
PACS numbers: 03.65.Ge.
\end{small}

\bigskip

\section{Introduction} 

"Do equations of motion determine quantum mechanical commutation relationships?" Wigner thought about this question in the middle of the twentieth century and investigated the harmonic oscillator problem in order to find an answer \cite{Wigner}. According to Wigner's analysis, the commutation relations may not always hold. Only a year later Yang showed that Wigner's conclusion is not exactly true. In \cite{Yang}, 
he demonstrated that with properly formulated conditions, i.e. a more rigid definition of appropriate Hilbert space and strict expansion theorem, the commutation relation would always hold. To illustrate his solution he defined a one dimensional reflection operator, $\hat{R}$,  with the action $\hat{R} \, f(x) = f(-x)$. Then, he introduced the following deformed Heisenberg algebra 
\begin{eqnarray}
\big[\hat{x},\hat{p}\big]=i\big(1+ \alpha \hat{R} \big),
\end{eqnarray}
with a Wigner parameter, $\alpha$, and a deformed momentum operator 
\begin{eqnarray}
\hat{p}= i \bigg( -\frac{d}{dx}+ \frac{\alpha}{2x} \hat{R} \bigg).
\end{eqnarray}
In 1987, Watanabe used these operators to solve one dimensional quantum harmonic oscillator \cite{Watanabe}. His findings presented a linear association between the Wigner parameter and ground-state energy state. Besides, he realized that the Hamiltonian operator has self-adjoint extension with the Wigner parameter. Meanwhile, Dunkl focused on the relation between differential-difference and reflection operators in pure mathematics point of view. In those years, Dunkl was focused on the relationship between reflection and differential-difference  operators from a purely mathematical perspective \cite{Dunkl0}. Just two years after Watanabe's paper, Dunkl introduced a novel combination of reflection and  differential-difference  operators of the form \cite{Dunkl1}
\begin{equation}\label{dunkloriginal}
D_x^{\alpha}=\frac{\partial}{\partial x}+\frac{\alpha}{x}(1-R).
\end{equation}
When the Dunkl derivative acts on a function, the results vary depending on whether the function has odd or even properties. In this respect there is a fundamental difference between the Dunkl and Yang operator \cite{ChungDunkl}. Therefore, in physics, the Dunkl derivative is used to take parity into account in the solution of the considered system \cite{G1, G2, G3, G4, G5, Ramirez1, Ramirez2, Sargol, Chung1, Ghaz, Mota1, Chungrev, Kim, Jang, Ojeda, Mota2, Mota3, Merad, ChungEPJP, Hassan, Dong, Bilel1, Bilel2, Seda, Mota2022}.

In literature, some generalizations of the Dunkl operator are being considered \cite{Chouchane, Karamov, Trim}. For example, Chouchane et al. proposed the following differential-difference operator \cite{Chouchane}
\begin{eqnarray}
D^{\alpha \beta}_x=\frac{d}{dx}+ \frac{(2\alpha+1)\coth x +(2\beta+1)\tanh x}{2}\big(1-R\big),
\end{eqnarray}
which is known as Jacobi-Dunkl operator \cite{Trim}. 
Last year, one of the author of the current manuscript, proposed a new generalized Dunkl operator \cite{ChungEPJP}. 
\begin{eqnarray}
D_{x}^{\alpha,\beta,\gamma}=\frac{\partial}{\partial x}+\frac{\alpha}{x}+\frac{\beta}{x}R+\gamma \frac{\partial}{\partial x}R,  \label{gDunkl}
\end{eqnarray}
with three parameters in one spatial dimension. One can easily observe that this generalization consists of new terms with  partial derivative and reflection operator. Therefore, this generalization is expected to lead to results that are more compatible with future experiments.  

Harmonic oscillator problem is one of the  fundamental and important phenomena of physics since the harmonic oscillator potential can be used as a model to approximate many physical phenomena quite well. Almost any system near equilibrium is assumed to act, at least approximately, as an harmonic oscillator since one can Taylor expand the potential energy and take the linear term as zero by construction. Therefore, the harmonic oscillator problem still is a hot topic since it applies to everything from atoms in a crystal to quantum fields. In the last decade Genest et al. studied isotropic Dunkl-oscillator model in the plane \cite{G1}. They constructed the Dunkl-oscillator Hamiltonian by employing the Dunkl derivative instead of the partial derivative. On the same year, they extended their work by considering anisotropic harmonic potential \cite{G2}. A year later, they handled the same models by algebraic methods \cite{G3}. Then, they considered the isotropic Dunkl-oscillator model in three spatial dimensions \cite{G4}. Relativistic Dunkl-oscillators are also been examined. For example, Sargolzaeipor et al. investigated the effects of the Wigner-Dunkl algebra on Dirac oscillator in \cite{Sargol}. One year later Mota et al. derived an exact solution of Dunkl-Dirac solution in the plane \cite{Mota1}. Last year, they presented a rigorous solution of two dimensional Dunkl-Klein Gordon oscillator  \cite{Mota2, Mota3}. Very recently, Hamil et al. gave the three dimensional solution of Dunkl-Klein Gordon oscillator in \cite{Bilel1}. Another interesting oscillator, namely the Dunkl-Duffin-Kemmer-Petiau oscillator, is investigated by Merad et al.  in the presense of the Dunkl derivative in \cite{Merad}.

There are many systems throughout the physical world that can be modeled by anharmonic oscillator. Basically, an anharmonic oscillator is a system that is not oscillating in harmonic motion. Therefore, anharmonicity is known to be the deviation from being a harmonic oscillator system and can be calculated using perturbation theory. Anharmonic potential models have been studied in various areas of physics with different methods such as modifying the harmonic oscillator's exact solution,  WKB technique, analytical continuation to Borel summability method, variational methods, etc... \cite{23,24,25,26,27,28,29,29b,30}.

Although we notice rigorous works on the solutions of the Dunkl-harmonic oscillator in literature, we see that Dunk-anharmonic oscillator solutions have not been found. Moreover, the most general form of the Dunkl-operator presents extended information regarding the parity of the considered system. With this motivation in this manuscript, we intend to examine the solutions of harmonic and anharmonic oscillators by using the generalized-Dunkl derivative instead of the usual partial derivative in the two-dimensional Klein-Gordon equation.

This manuscript is organized as follows: In section 2 we introduce the generalized-Dunkl Laplacian in two dimensions and map the operator to polar coordinates. After we find the azimuthal and radial parts, in section 3, we investigate the generalized-Dunkl harmonic oscillator solutions with Nikiforov-Uvarov (NU) method. Then, in section 4 we consider the generalized-Dunkl anharmonic potential instead of harmonic one and obtain a solution by Quasi-exact solvability (QES) method. Finally, we conclude the manuscript in section 5.

\section{Two dimensional generalized-Dunkl Laplacian \label{sec2}}

In this section, we construct the mathematical base of the manuscript by constructing the generalized-Dunkl Laplacian operator by employing the generalized-Dunkl derivative given in Eq. \eqref{gDunkl}. In two-dimension each derivatives carries three Wigner parameters, and the square of generalized-Dunkl derivatives become:
\begin{subequations}
\begin{align}
(D_x^{\alpha_1,\beta_1,\gamma_1})^2 & =(1-\gamma_1^2)\frac{\partial^2}{\partial x^2}+\frac{2}{x}(\alpha_1-\beta_1\gamma_1)\frac{\partial}{\partial x}+\frac{\alpha_1^2-\beta_1^2-\alpha+\beta_1\gamma_1}{x^2}-\frac{\beta_1-\alpha_1\gamma_1}{x^2}R_1,\\
(D_y^{\alpha_2,\beta_2,\gamma_2})^2 & =(1-\gamma_2^2)\frac{\partial^2}{\partial y^2}+\frac{2}{y}(\alpha_2-\beta_2\gamma_2)\frac{\partial}{\partial y}+\frac{\alpha_2^2-\beta_2^2-\alpha+\beta_2\gamma_2}{y^2}-\frac{\beta_2-\alpha_2\gamma_2}{y^2}R_2,
\end{align}
\end{subequations} 
where two reflection operators satisfy 
\begin{equation}
R_x\ f(x,y)=f(-x,y),\quad \quad R_y\ f(x,y)=f(x,-y).
\end{equation}
Applying new variables
\begin{subequations} \label{polcor}
\begin{eqnarray}
x&=& \sqrt{1-\gamma_1^2}\rho \cos\phi, \\  y&=&\sqrt{1-\gamma_2^2}\rho \sin\phi, 
\end{eqnarray}
\end{subequations}
with the constraint $-1<\gamma_i<1$, we find the generalized-Dun1kl Laplacian in polar coordinates as follows:
\begin{equation} 
\Delta_D\equiv(D_x^{\alpha_1,\beta_1,\gamma_1})^2+(D_y^{\alpha_2,\beta_2,\gamma_2})^2=\frac{\partial^2}{\partial\rho^2}+\frac{2(\frac{1}{2}+\xi_1+\xi_2)}{\rho}\frac{\partial}{\partial\rho}+\frac{B_\phi}{\rho^2}. \label{gLappol}
\end{equation}
Here $B_\phi$ operator reads 
\begin{equation}
B_\phi=\frac{\partial^2}{\partial\phi^2}+2(\xi_2\cot\phi-\xi_1\tan\phi)\frac{\partial}{\partial\phi}+\frac{\mu_1+\nu_1R_1}{\cos^2\phi}+\frac{\mu_2+\nu_2R_2}{\sin^2\phi},
\end{equation}
with
\begin{subequations} \label{Wp}
\begin{align}
\xi_1&=\frac{(\alpha_1-\beta_1\gamma_1)}{(1-\gamma_1^2)},   &    \xi_2&=\frac{(\alpha_2-\beta_2\gamma_2)}{(1-\gamma_2^2)},\\
\mu_1&=\frac{(\alpha_1^2-\beta_1^2-\alpha_1+\beta_1\gamma_1)}{(1-\gamma_1^2)},   &   \mu_2&=\frac{(\alpha_2^2-\beta_2^2-\alpha_2+\beta_2\gamma_2)}{(1-\gamma_2^2)}, \\
\nu_1&=-\frac{(\beta_1-\alpha_1\gamma_1)}{(1-\gamma_1^2)},   &   \nu_2&=-\frac{(\beta_2-\alpha_2\gamma_2)}{(1-\gamma_2^2)}.
\end{align}
\end{subequations}
If one takes all the Wigner parameter as zero, then the parameters given in Eq. \eqref{Wp} disappear and $B_\phi$ operator reduces to a simple form, $\frac{\partial^2}{\partial\phi^2}$.
Since $B_\phi$ does not depend on radial variable, we can assume an eigenfunction as $\psi(\rho,\phi)=R(\rho)\Phi(\phi)$. Then, the azimuthal part leads to
\begin{equation}\label{e11}
B_\phi \Phi(\phi)=-m'^{\,2}\Phi(\phi).
\end{equation}
By defining a new variable, $u=\cos^2\phi$, Eq. \eqref{e11} becomes
\begin{equation}\label{e5}
\begin{split}
&\Phi''(u)+\left[\frac{(\frac{1}{2}+\xi_1)-(1+\xi_1+\xi_2)u}{u(1-u)}\right]\Phi'(u)\\
+&\left[\frac{(\mu_{1}+\nu_{1} R_{1})+(\mu_{2}+\nu_{2}R_{2}-\mu_{1}-\nu_{1}R_{1}+m'^{\,2})u-m'^{\,2} u^2}{4u^2(1-u)^2}\right]\Phi(u)=0.
\end{split}
\end{equation}
In order to solve Eq. \eqref{e5} we use NU method \cite{31,32,33,34,35}. We obtain eigenfunction in the form of
\begin{equation}
	\Phi(\phi)=e^{-\frac{\cos^2\phi}{4}(2-2\xi_1-2\xi_2+k_1+k_2)}\cos\phi^{\frac{1-2\xi_1+k_1}{2}}L_{n_\phi}^{\frac{k_1}{2}}(\frac{4+k_1+k_2}{2}\cos^2\phi),
\end{equation}
with
\begin{equation}
	k_i=\sqrt{(1-2\xi_i)^2-4\mu_i-4\nu_iR_i}, \quad (i=1,2).
\end{equation}
Here, $L_n^r$ denotes the generalized Laguerre polynomial, and $m'$ satisfies the following relation
\begin{equation}
\begin{split}
m'^2=\, & 2n_\phi(2n_\phi+2+k_1+k_2)+\frac{3}{2}-(\mu_1+\mu_2)+(k_1+k_2+\frac{k_1k_2}{2})-(\xi_1+\xi_2+2\xi_1\xi_2)\\
-&(\nu_1R_1+\nu_2R_2).
\end{split}
\end{equation}
We observe that in the usual case, we get $k_1=k_2=1$ and modified quantum numbers produce only integer values. After obtaining the eigenvalues of the azimuthal part we are ready to solve the radial part. In the following two sections, we will present radial part solutions of the Klein-Gordon equation with harmonic and anharmonic potential energies in the presence of generalized Dunkl derivative, respectively.

\section{Harmonic oscillator case\label{Ha}}
The Klein-Gordon equation is known to be the most appropriate relativistic wave equation to describe the dynamics of spin-zero particles. In the natural units, $\hbar=c=1$, with the presence of scalar and vector potential energies it takes the following form \cite{BCL}
\begin{equation}\label{eq6}
	\left[\nabla^2+(E-V(|\Vec{r}|))^2-(m+S(|\Vec{r}|))^2\right]\psi(|\Vec{r}|)=0.
\end{equation}
Let us assume that the vector and scalar potential energy functions have the following same magnitude \begin{equation}
V=\frac{1}{2}mw^2(x^2+y^2),
\end{equation}
which transform to 
\begin{equation}
V(\rho)=\frac{1}{2}mw^2(1-a^2)\rho^2,
\end{equation}
according to the defined polar coordinates in Eq. \eqref{polcor} with $\gamma_1=\gamma_2=a$. With the use of generalized-Dunkl Laplacian in Eq.\eqref{eq6}, the generalized-Dunkl Klein-Gordon equation reads
\begin{eqnarray}
\frac{\partial^2\psi(\rho)}{\partial\rho^2}+\frac{2}{\rho}\left(\frac{1}{2}+\xi_1+\xi_2\right)\frac{\partial\psi(\rho)}{\partial\rho}
+\left(E^2-\frac{m'^{\,2}}{\rho^2}-2EV(\rho)-m^2-2mV(\rho)\right)\psi(\rho)=0. \label{gDKG}
\end{eqnarray}
By introducing a new variable, $s=\rho^2$, we express Eq. \eqref{gDKG} in the form of
\small
\begin{eqnarray}
s^2\frac{d^2\psi}{ds^2}+s\left(1+\xi_1+\xi_2\right)\frac{d\psi}{ds}
+\bigg[s^2\left(\frac{m\omega (E+m) (a^2-1)}{4}\right)+s\left(\frac{E^2-m^2}{4}\right)-\frac{m'^{\,2}}{4}\bigg]\psi_{n,m^\prime}(s)=0.
\end{eqnarray}
\normalsize
Using the NU method, we obtain the wavefunction 
\small
\begin{eqnarray}
\psi_{n,m'}(s) =s^{-\frac{\xi_1+\xi_2}{2}+\sqrt{\frac{1}{4}(\xi_1+\xi_2)^2+\frac{m'^{\,2}}{4}}}  e^{-\sqrt{\frac{m\omega (E+m) (a^2-1)}{4}}} L^{\sqrt{(\xi_1+\xi_2)^2+m'^{\,2}}}_{n}(\sqrt{m\omega (E+m) (a^2-1)}s), \label{ss}
\end{eqnarray}
\normalsize
where the energy eigenvalue function reads
\begin{eqnarray}
E_{n,m'} &=&\frac{1}{3}\left[m+\frac{2m^2}{\mathcal{M}^{\frac{1}{3}}}+2\mathcal{M}\right]^{\frac{1}{3}},
\end{eqnarray}
with
\begin{eqnarray}
\mathcal{M}&=&m\left[3\sqrt{3}M^2\omega^2\sqrt{(a^2-1)(2m^2+27(a^2-1))}-(m^2+27(a^2-1)M^2w^2)\right], \,\,\,\,\,\,\\
M&=&\frac{1}{2}+n+\frac{1}{2}\sqrt{m'^{\,2}+{(\xi_1+\xi_2)}^2}.
\end{eqnarray}
In Fig. 1 we have plotted energy versus $a$. we observe that when $a$ increases the energy eigenvalue decreases. When $a$ goes to $1$, the energy becomes independent of the quantum number $n$, degeneracy increases. Fig. 1 also presents the energy spectrum for positive and negative parities. Obviously, the energy spectrum differs for positive and negative parities and with increasing $a$ the energy between two states will decrease.
\begin{figure}[htbp]
\centering
\includegraphics[width=12cm]{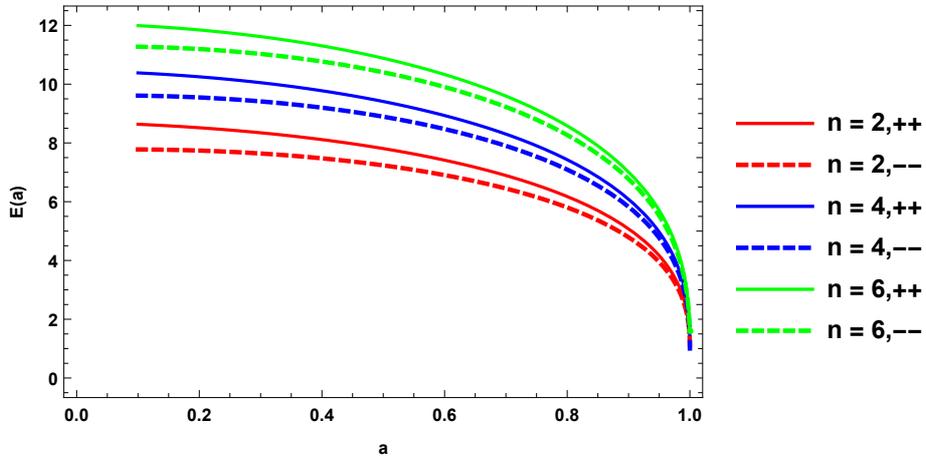}
\caption{Plot of energy eigenvalues versus a for different n $n_\phi=2,\alpha_1=\alpha_2=0.5,\beta_1=\beta_2=0.5,R_1=R_2=+1,R_1=R_2=-1,m=w=1$.}
\end{figure}

We also demonstrate some of the wave eigenfunctions for different values of quantum number, $n$,  in Fig. 2.  
\begin{figure}[htbp]
\centering
\includegraphics[width=12cm]{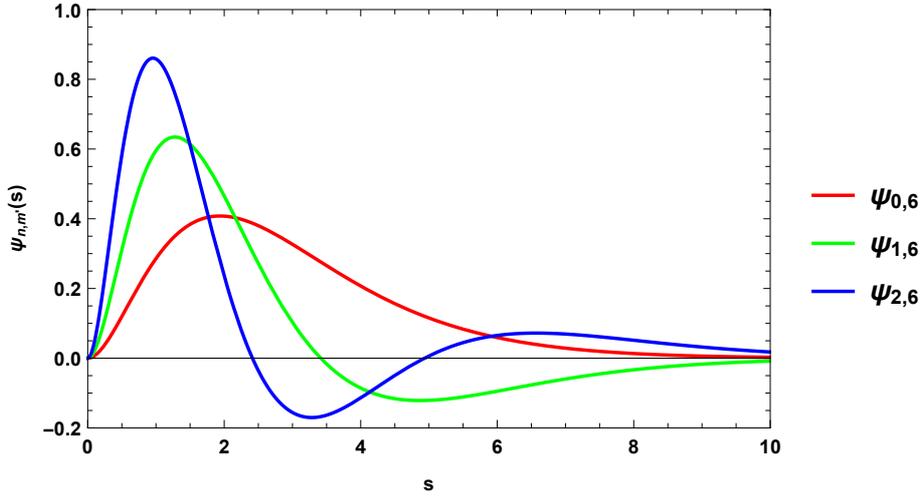}
\caption{Ground and first two excited positive parity eigenstates for $n_\phi=m=w=1$, $a=-0.6$, and $\alpha_1=\alpha_2=\beta_1=\beta_2=0.5$. }
\end{figure}

\newpage
\section{Anharmonic oscillator case}
In this section, we attempt to obtain the solution of the following anharmonic potential 
\begin{equation}
	V(x,y)=\Omega(x^2+y^2)+\Lambda{(x^2+y^2)}^2+\Gamma{(x^2+y^2)}^3.
\end{equation}
in Eq. \eqref{gDKG}. Following the previous choice, $\gamma_1=\gamma_2=a$, we define new parameters
\begin{subequations}
\begin{eqnarray}
     w&\equiv&(1-a^2)\Omega, \\
  \lambda&\equiv&{(1-a^2)}^2\Lambda,\\
  \eta&\equiv&{(1-a^2)}^3\Gamma,
\end{eqnarray}
\end{subequations}
to express the potential energy as
\begin{equation}
	V(\rho)=w\rho^2+\lambda\rho^4+\eta\rho^6.
\end{equation}
Then, we express Eq. \eqref{gDKG} by using the anharmonic potential
\begin{eqnarray}
\left[\frac{\partial^2}{\partial\rho^2}+\frac{1+2(\xi_1+\xi_2)}{\rho}\frac{\partial}{\partial\rho}-\frac{m'^{\,2}}{\rho^2}
+E^2-m^2-2{(w\rho^2+\lambda\rho^4+\eta\rho^6)}{(E+m)}\right]\psi_{n,m^\prime}(\rho)=0. \label{ss2}
\end{eqnarray}
We define a new variable, $Z\equiv\rho^2$, Eq. \eqref{ss2} takes the form
\begin{eqnarray} 
\left[\frac{d^2}{dZ^2}+\frac{(1+\xi_1+\xi_2)}{Z}\frac{d}{dZ}-\frac{m'^{\,2}}{4Z^2}+\frac{E^2-m^2}{4Z}-\frac{(w+\lambda Z+\eta Z^2)(E+m)}{2}\right]\psi_{n,m'}(Z)=0. \label{e7}
\end{eqnarray}
We make an ansatz
\begin{equation}
	\psi_{n,m^\prime}(Z)=Z^{\frac{-\xi_1-\xi_2}{2}}U_{n,m^\prime}(Z).
\end{equation}
Then, Eq. \eqref{e7} becomes
\begin{equation}
\left[{\frac{d^2}{dZ^2}+\frac{1}{Z}\frac{d}{dZ}+\frac{\lambda_1}{Z^2}+\frac{\lambda_2}{Z}+\lambda_3+\lambda_4 Z+\lambda_5Z^2}\right]U_{n,m'}(Z)=0.
\end{equation}
Here,
\begin{subequations}
\begin{align}
\lambda_1 &=-\frac{m'^{\,2}+{(\xi_1+\xi_2)}^2}{4},\\
\lambda_2 &=\frac{(E-m)(E+m)}{4},\\
\lambda_3 &=-\frac{w(E+m)}{2},\\
\lambda_4 &=-\frac{\lambda(E+m)}{2},\\
\lambda_5 &=-\frac{\eta(E+m)}{2}.
\end{align}
\end{subequations}
We notice that Eq. \eqref{e7} has the form of $ H U_{n,m'}=0$, where the Hamilton operator is
\begin{equation}
H=\frac{d^2}{dZ^2}+\frac{1}{Z}\frac{d}{dZ}+\frac{\lambda_1}{Z^2}+\frac{\lambda_2}{Z}+\lambda_3+\lambda_4 Z+\lambda_5Z^2.
\end{equation}
For the solution, we employ the QES technique  which is based on gauge transformation \cite{36,37,38,39,40,41,42,43,44,45,46,47,48},
\begin{equation}
	\tilde{H}\tilde{U}_{n,m'}=0.
\end{equation}
The gauge operator, $G$, is introduced to  transform the Hamiltonian operator and the wave function as $\tilde{H}=G^{-1}.H.G$, and $	U_{n,m^\prime}=G.\tilde{U}_{n,m^\prime}$, respectively. Then, the transformed wave function takes the form of
\begin{equation}
\tilde{U}_{n,m^\prime}=\sum_{k=1}^{\infty}a_kZ^k,\quad\quad n=1,2,3,...
\end{equation}
All QES differential equations transform under the generators
\begin{subequations}\label{eq12}
\begin{align}
J^+_n & =Z^2\frac{d}{dZ}-nZ,\\
J^0_n & =Z\frac{d}{dZ}-\frac{n}{2},\\
J^-_n & =\frac{d}{dZ},
\end{align}
\end{subequations}
that satisfy the commutation relations of Sl(2) Lie algebra
\begin{subequations}
\begin{align}
\left[J^{\mp},J^0\right] & =\pm J^\pm, \\
\left[J^+,J^-\right] & =-2J^0.
\end{align}
\end{subequations}
Therefore, the transformed Hamiltonian can be written in terms of these generators 
\begin{eqnarray}
\tilde{H} & =& C_{++}J^+_n J^+_n + C_{+0} J^+_n J^0_n +  C_{+-} J^+_n J^-_n+ C_{0-} J^0_n J^-_n +
		 C_{--} J^-_n J^-_n+ C_{+} J^+_n \nonumber \\ &&+C_0J^0_n+C_-J_n^-+C. \label{eq14}
\end{eqnarray}
In principle any QES Hamiltonian can be transformed to following equation 
\begin{eqnarray}
\tilde{H} & =p_4\frac{d^2}{dZ^2}+p_3\frac{d}{dZ}+p_2, \label{Eq151}
\end{eqnarray}
where $P_i$ are the polynomials of degree $i$
\begin{subequations} \label{Eq152}
\begin{eqnarray}
p_4 & =&C_{++}Z^4+C_{+0}Z^3+C_{+-}Z^2+C_{0-}Z+C_{--},\\
p_3 & =&C_{++}(2-2n)Z^3+\Big(C_++C_{+0}\big(1-\frac{3n}{2}\big)\Big)Z^2+(C_0-nC_{+-})Z+(C_--\frac{n}{2}C_{0-}),\,\,\,\,\,\,\,\,\,\,\,\,\\
p_2 &=&C_{++}n(n-1)Z^2+(\frac{n^2}{2}C_{+0}-nC_+)Z+(C-\frac{n}{2}C_0).
\end{eqnarray}
\end{subequations}
Then, we propose the following transformation,
\begin{equation}
	U_{n,m^\prime}(Z)=Z^A e^{-BZ^2-DZ}\tilde{U}_{n,m^\prime}(Z),
\end{equation}
so the transformed Hamiltonian reads
\begin{equation}\label{eq17}
	\tilde{H}=\frac{d^2}{dZ^2}+\left(\frac{2A+1}{Z}-4BZ-2D\right)\frac{d}{dZ}+\frac{\lambda_2-2AD-D}{Z}+\lambda_3-4AB-4B+D^2,
\end{equation}
where
	\begin{eqnarray}
		A=\sqrt{-\lambda_1},\quad\quad
		B^2=-\frac{\lambda_5}{4},\quad\quad
		D =\frac{-\lambda_4}{4B}.
	\end{eqnarray}
By matching Eq. \eqref{Eq151} and Eq. \eqref{eq17}, we get
\begin{equation}\label{eq19}
\begin{split}
C_{++} &=0, \hspace{1cm} C_{+0}=0, \hspace{1cm} C_{+-}=0, \hspace{1cm} C_{0-}=1, \hspace{1cm} C_-=\frac{n}{2}+2A+1, \\
C_{--} &=0,\hspace{1cm} C_{+}=-4B, \hspace{0.6cm} C_{0}=-2D, \hspace{0.5cm} C=-nD-2AD-D+\lambda_2,\\
-n C_{+} &=\lambda_3-4AB-4B+D^.
\end{split}
\end{equation}
Then, by using $-nC_+$ we obtain energy eigenvalue function 
\begin{equation}
\begin{split}
E_{n,m\prime} 
 &=\frac{-m\lambda^4+32\eta^3\left(m'^2+4(1+n)^2+{(\xi_1+\xi_2)}^2+
4\sqrt{m'^2+{(\xi_1+\xi_2)}^2}(1+n)\right)}{(\lambda^2-4\eta w)^2}\\
& +\frac{8m\eta w (\lambda^2 -2\eta w)} {(\lambda^2-4\eta w)^2}.
\end{split}
\end{equation}
In Fig. 3, we depict energy eigenvalue functions versus $a$ for different values of $n$. We observe that when $a$ goes to $1$, the energy eigenfunction  becomes independent from the quantum number $n$ and all states are degenerate. The behavior of energy spectra for parity positive and negative is similar to the previous section's results: An increase in $a$ causes a decrease in the   difference of energy  between two states.
\begin{figure}[htb]
\centering
\includegraphics[width=12cm]{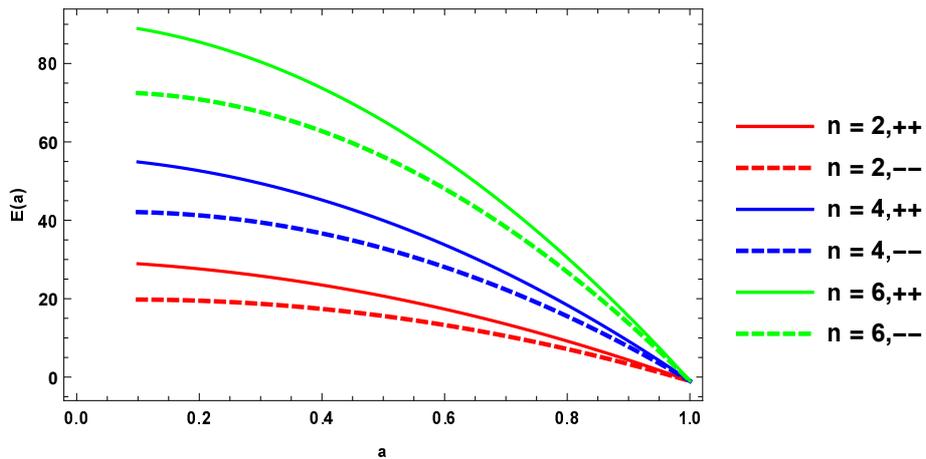}
\caption{Plot of energy eigenvalues versus $a$ for different $n$ and $n_\phi=2,\alpha_1=\alpha_2=0.5,\beta_1=\beta_2=0.5,m=1, \Gamma=2, \Omega=4, \Lambda=0.5$. }
\end{figure}

By substituting the coefficients given in Eq. \eqref{eq19} into Eq. \eqref{eq14}, we obtain
\begin{equation}
\tilde{H}=J^0_nJ^-_n-4BJ^+_n-2DJ^0_n+(2A+1+\frac{n}{2})J^-_n-nD+\lambda_2-2AD-D
\end{equation}
Then, we use the generators and express the transformed Hamiltonian as
\begin{equation}
\tilde{H}=Z\frac{d^2}{dZ^2}+(-4BZ^2-2DZ+2A+1)\frac{d}{dZ}+4nBZ+\lambda_2-2AD-D.
\end{equation}
By using $\tilde{H}\tilde{U}_{n,m\prime}=0$ , we obtaine a recursion relation 
\begin{equation}
a_{k+1}=-\frac{-2DA+\lambda_2-D(2k+1)}{(k+1)(2A+1+k)}a_{k}+\frac{-8Bk}{(k+1)(2A+1+K)}a_{k-1},
\end{equation}
where
\begin{subequations}
\begin{align}
&a_0=1,\\
&a_1=\frac{-(\lambda_2-2AD-D)}{(2A+1)}a_0,\\
&a_2=\frac{-8B}{4(A+1)}a_0-\frac{(\lambda_2-2AD-3D)}{4(A+1)}a_1.
\end{align}
	\end{subequations}
Then, the  first excited wave eigenfunction state stands as follows:
\begin{equation}
U_{1,m^\prime}=Z^Ae^{-BZ^2-DZ}\left(1+\frac{2AD-\lambda_2+D}{2A+1}Z\right).
\end{equation}
In Fig. 4, we plot ground state and first excited state of wave function in terms of $Z$.
\begin{figure}[htb]
	\centering
	\includegraphics[width=12cm]{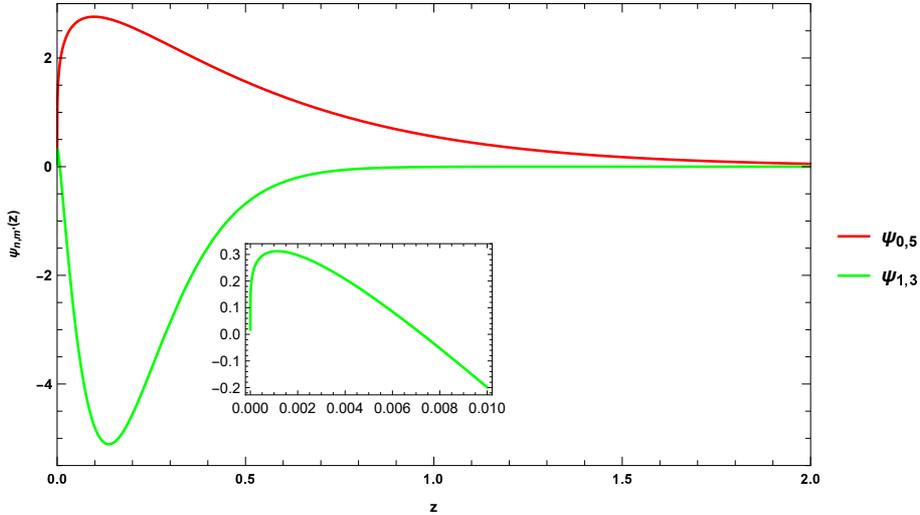}
	\caption{Plot of energy eigenvalues versus a for $(n=0, n_\phi=1,\alpha_1=0.5,\alpha_2=7.896152,\beta_1=\beta_2=1,R_1=R_2=1,m=1, \Omega=1, \Lambda=10, \Gamma=1, a=-0.8), (n=1, n_\phi=1,\alpha_1=0.5,\alpha_2=6.908326,\beta_1=\beta_2=1,R_1=R_2=1,m=1, \Omega=0.2, \Lambda=3, \Gamma=1, a=-0.5)$.}
\end{figure}

\section{Conclusion\label{Conc}}
In the present manuscript, we solved two dimensional Klein-Gordon equation for harmonic and anharmonic potential energies in the presence of generalized-Dunkl derivative formalism. The latter Dunkl derivative has three Wigner parameters and is substituted for the ordinary partial derivative in the presence of vector and scalar potential couplings. After we expressed the generalized-Dunkl Laplacian, we mapped mapped it to polar coordinates. We observed that the latter form has azimuthal and radial parts. We used the NU method to solve the azimuthal part and obtained a modified quantum number that can take any real values. We observed that the latter quantum number produces integer values in the usual case. Next, for the radial part we considered first the harmonic, and then, the anharmonic potential energies. We employed the NU and QES methods to derive solutions, respectively. In the harmonic oscillator case, we found energy eigenvalue function and their corresponding eigenstates. We observed that when the Wigner parameters $\gamma_i$ go to $1$ all the states become degenerate. Then, we considered an anharmonic potential in the sextic-potential energy form.  We found that in the anharmonic case, the differential equation is not convertible to the hypergeometric differential equation. Therefore, we used QES method to obtain solution. In so doing, we use series solution to determine eigenstate functions. We observed that the energy eigenvalue function behaves similar to the harmonic potential case. 

\section*{Acknowledgments}

One of the authors of this manuscript, BCL, is
supported by the Internal Project, [2022/2218], of Excellent Research of the
Faculty of Science of Hradec Kr\'alov\'e University.








\begin{thebibliography}{99}

%
    \bibitem{Wigner}E. P. Wigner, Phys. Rev. {\bf 77}, 711 (1950).
%
    \bibitem{Yang}L. M. Yang, Phys. Rev. {\bf 84}, 788 (1951).
%
    \bibitem{Watanabe}
    S. Watanabe, J. Math. Phys. {\bf 30}(2), 376 (1989). 
%
    \bibitem{Dunkl0}C. F. Dunkl, Math. Z. {\bf 197}, 33 (1988).
%
    \bibitem{Dunkl1}C. F. Dunkl, T. Am. Math. Soc. {\bf 311}(1), 167 (1989).
%
    \bibitem{ChungDunkl} W. S. Chung, H. Hassanabadi, Mod. Phys. Lett. A \textbf{36}(18), 2150127 (2021). 
%
%
\bibitem{G1} V. Genest, M. Ismail, L. Vinet, A. Zhedanov, J. Phys. A \textbf{46}, 145201 (2013).
%
\bibitem{G2} V. Genest,  L. Vinet, A. Zhedanov, J. Phys. A \textbf{46}, 325201 (2013).
%
\bibitem{G3} V. Genest, M. Ismail, L. Vinet, A. Zhedanov, Commun. Math. Phys. \textbf{329}, 999 (2014).
%
\bibitem{G4} V. Genest, L. Vinet, A. Zhedanov, J. Phys. Conf. Ser. \textbf{512}, 012010 (2014).
%
\bibitem{G5} V. Genest, A. Lapointe, L. Vinet, Phys. Lett. A \textbf{379}, 923 (2015).
%
\bibitem{Jang} E. J. Jan, S. Park, W. S. Chung, J. Kor. Phys. Soc. \textbf{68}(3), 379 (2016).
%
\bibitem{Ramirez1} M. Salazar-Ramirez, D. Ojeda-Guill\'en, V. D. Granados, Eur. Phys. J. Plus \textbf{132}, 39 (2017).
%
\bibitem{Ramirez2} M. Salazar-Ramirez, D. Ojeda-Guill\'en, R. D. Mota, V. D. Granados, Mod. Phys. Lett. A \textbf{33}(20), 1850112 (2018).
%
\bibitem{Sargol} S. Sargolzaeipor, H. Hassanabadi, W. S. Chung, Mod. Phys. Lett. A \textbf{33}(25), 1850146 (2018).
%
\bibitem{Chung1} W. S. Chung, H. Hassanabadi, Mod. Phys. Lett. A \textbf{34}(24), 1950190 (2019).
%
\bibitem{Ghaz} S. Ghazouani, I. Sboui, M. A. Amdouni, M. B. El Hadj Rhouma, J. Phys. A: Math. Theor. \textbf{52}, 225202 (2019).
%
\bibitem{Mota1} R. D. Mota, D. Ojeda-Guill\'en, M. Salazar-Ram\'irez, V. D. Granados, Ann. Phys. \textbf{411}, 167964 (2019).
%
\bibitem{Chungrev} W. S. Chung, H. Hassanabadi, Rev. Mex. Fis.  \textbf{66}(3), 308 (2020).
%
\bibitem{Kim} Y. Kim, W. S. Chung, H. Hassanabadi, Rev. Mex. Fis.  \textbf{66}(4), 411 (2020).
%
\bibitem{Ojeda}  D. Ojeda-Guill\'en, R. D. Mota, M. Salazar-Ram\'irez, V. D. Granados, Mod. Phys. Lett. A \textbf{35}(31), 2050255 (2020).
%
\bibitem{Mota2} R. D. Mota, D. Ojeda-Guill\'en, M. Salazar-Ram\'irez, V. D. Granados, Mod. Phys. Lett. A \textbf{36}(10), 2150066 (2021).
%
\bibitem{Mota3} R. D. Mota, D. Ojeda-Guill\'en, M. Salazar-Ram\'irez, V. D. Granados, Mod. Phys. Lett. A \textbf{36}(23), 2150171 (2021).
%
\bibitem{Merad} A. Merad, M. Merad, Few-Body Syst. \textbf{62}, 98 (2021).
%
\bibitem{ChungEPJP} W. S. Chung, H. Hassanabadi, Eur. Phys. J. Plus.  \textbf{136}, 239 (2021).
%
\bibitem{Hassan} H. Hassanabadi, M. de Montigny, W. S. Chung, Physica A  \textbf{580}, 126154 (2021).
%
\bibitem{Dong}S. H. Dong, W. H. Huang, W. S. Chung, P. Sedaghatnia, H. Hassanabadi, EPL \textbf{135}, 30006 (2021).
%
\bibitem{Bilel1} B. Hamil, B. C. L\"utf\"uoğlu, arXiv:2112.09948 [quant-ph].
%
\bibitem{Bilel2} B. Hamil, B. C. L\"utf\"uoğlu, Eur. Phys. J Plus, \textbf{137}, 812 (2022). 
%
\bibitem{Seda} P. Sedaghatnia, H. Hassanabadi, W. S. Chung, B. C. L\"utf\"uoğlu, J. Kříž, \textit{Under review} (2022).
%
\bibitem{Mota2022} R. D. Mota, D. Ojeda-Guill\'en, Mod. Phys. Lett. A \textbf{37}(01), 2250006 (2022).
%
\bibitem{Chouchane} F. Chouchane, M. Mili,  K. Trim\'eche, J. Analy. Appl. \textbf{01}(04), 387 (2003).
%
\bibitem{Karamov} I. I. Karamov, V. V. Napalkov, Ufa Math. J. \textbf{6}(1), 56 (2014). 
%
\bibitem{Trim} K. Trimèche,  Mediterr. J. Math. \textbf{12}, 349 (2015).
%
\bibitem{23}
C. M. Bender, T. T. Wu, Phys. Rev. \textbf{184}, 1231 (1969). 
%
\bibitem{24}
 S. Graffi, V. Gricchi, B. Simon. Phys. Lett. B \textbf{32}(7), 631 (1970). 
%
\bibitem{25}
 V. Singh, S. N. Biswas, K. Datta. Phys. Rev. D. \textbf{18}, 1901 (1978). 
%
\bibitem{26} P. B. Abraham, H. E. Moses
Phys. Rev. A \textbf{22}, 1333  (1980); Erratum Phys. Rev. A \textbf{23}, 2088 (1981).
%
\bibitem{27} M. Znojil, Phys. Rev. D \textbf{24}(4), 903 (1981).
%
\bibitem{28}P. Buganu, R. Budaca, J. Phys. G: Nucl. Part. Phys. \textbf{42}, 105106 (2015). 
%
\bibitem{29} B. M. Villegas-Martínez, H. M. Moya-Cessa, F. Soto-Eguibar,  Eur. Phys. J. D \textbf{74}, 137 (2020). 
%
\bibitem{29b} A. V. Turbiner, J. C. del Valle, J. Phys. A: Math. Theor. \textbf{54}, 295204 (2021).
%
\bibitem{30} D. Bambusi, B. Langella, M. Rouveyrol, Commun. Math. Phys. \textbf{390}, 309 (2022).

%
\bibitem{31}
A. F. Nikiforov, V.B. Uvarov.  Special Functions of Mathematical Physics. Birkhauser, Basel,(1988).
%
  \bibitem{32}
 B. Gönül, K. Köksal. Phys. Scr. \textbf{75}, 686 (2007).
%
 \bibitem{33}
 C. Tezcan, R. Sever, Int. J. Theor. Phys. \textbf{48}, 337 (2009). 
%
\bibitem{34}
 O. Aydoğdu, R. Sever, Phys. Scr. \textbf{80}, 015001 (2009).
%
 \bibitem{35}
 H. Hassanabadi, E, Maghsoodi, S, Zarrinkamar, H. Rahimov, J. Math. Phys. \textbf{53}, 022104 (2012).
%
\bibitem{BCL} B. C. Lütfüoğlu, Eur. Phys. J. Plus \textbf{133}, 309 (2018). 
%
 \bibitem{36}
 A. V. Turbiner, Commun. Math. Phys. \textbf{118}, 467 (1988).
 \bibitem{37}  M. Znojil, J. Phys. A: Math. Gen. \textbf{27}, 7491 (1994).
%
 \bibitem{38}  M. Znojil, Phys. Lett. A \textbf{359}, 21 (2006).
%
\bibitem{39}
C. L. Ho, Annals Phys. \textbf{321}, 2170 (2006).
%
\bibitem{40}
R. Sasaki, J. Math. Phys. \textbf{48}, 122105 (2007).
%
\bibitem{41}
R. Sasaki, W. L. Yang, Y. Z. Zhang, SIGMA, \textbf{5}, 104 (2009).
%
\bibitem{42}
Y. H. Lee, J. Links, Y. Z. Zhang, J. Phys. A Math. Theor. \textbf{44}, 482001 (2011).
%
\bibitem{43}
H. Panahi, M. Baradaran,  Mod. Phys. Lett. A \textbf{27}(30), 1250176 (2012).
%
\bibitem{44} M. Znojil, Phys. Lett. A \textbf{380}, 1414 (2016).
%
 \bibitem{45}
 A. V. Turbiner, Phys. Rept. \textbf{642}, 1 (2016).
%
 \bibitem{46}
 C. Quesne, Eur. Phys. J. Plus \textbf{132}, 450 (2017).
%
 \bibitem{47} C. Quesne,  J. Phys. Conf. Ser. \textbf{1071}, 012016 (2018).
%
\bibitem{48}
B. C. L\"{u}tf\"{u}o\u{g}lu, J. K\v{r}\'{i}\v{z}, P. Sedaghatnia, H. Hassanabadi, Eur. Phys. J. Plus \textbf{135}, 691 (2020).
 
\end{thebibliography}
\end{document}